\documentclass[journal=jceda8,manuscript=article,layout=twocolumn]{achemso}
\usepackage{caption}
\usepackage[version=3]{mhchem} 
\usepackage{booktabs}  
\usepackage{cuted}
\usepackage{xr}

\makeatletter

\newcommand*\datastatementname{Data Availability Statement}
\makeatother

\author{Titus S. van Erp}
\email{titus.van.erp@ntnu.no}%
\affiliation{Department of Chemistry and Biomedical Science, Norwegian University of Science and Technology, Trondheim, Norway}

\title[Eyring vs Arrhenius]{Is the Eyring Plot Misleading? A Case for Arrhenius Analysis of Activation Parameters}

\abbreviations{TST}

\begin{document}
\begin{strip}
\vspace{-2cm} 
\begin{abstract}
A common view in physical chemistry literature is that the Eyring representation, in which a linear fit of $\ln(k/T)$ versus $1/T$ is attempted, is more fundamental than the Arrhenius representation, $\ln(k)$ versus $1/T$. This perception is typically motivated by its derivation from statistical mechanics and quantum mechanics, and by the interpretation of the intercept in terms of the activation entropy $\Delta S^\ddagger$, whereas the Arrhenius equation and its prefactor are often regarded as purely phenomenological. However, harmonic approximation models yield exact linearity in Arrhenius plots but not in Eyring plots, although for real experimental data both generally appear equally linear within typical experimental accuracy. Furthermore, the impression that the Eyring formulation is inherently quantum mechanical arises from the presence of the Planck constant in the prefactor, whereas this term results from normalization conventions in the partition function. This also highlights an interpretational issue in $\Delta G^\ddagger$, which is based on partition functions of different dimensionality between reactant and transition state. This dimensional mismatch can be reformulated in an alternative representation that improves interpretability and reduces to an Arrhenius-type expression in which the prefactor is directly related to an entropy of activation. In this framework, both activation enthalpy and entropy obtained from an Arrhenius fit are arguably more physically relevant than the corresponding Eyring fit values.
\end{abstract}
\end{strip}

\section{Introduction}
\label{sec:introduction}
Several expressions have been proposed in the literature to describe chemical reaction rate constants as a function of temperature within the framework of transition state theory (TST). Among the most commonly encountered are the Arrhenius equation,\cite{Arrhenius1889} the Eyring equation,\cite{Eyring1935} and Chandler's reactive flux expression:\cite{Chandler1978}
\begin{align}
   \label{eq:3k}
   k &= A \, e^{-\beta E_a} 
   \quad \text{(Arrhenius equation)}\\[4pt]
   k &= \frac{k_B T}{h} \, e^{-\beta \Delta G^\ddagger}  
   \quad \text{(Eyring equation)} \nonumber \\[4pt]
   k &= \sqrt{\frac{k_B T}{2 \pi m}_q} \,
   \frac{e^{-\beta G(q^\ddagger)}}{\int_{-\infty}^{q^\ddagger} e^{-\beta G(q)} \, dq}
   \quad \text{(reactive flux)} \nonumber
\end{align}
where $k_B$ and $h$ are the Boltzmann and Planck constants, $T$ the temperature, and $\beta = 1/(k_B T)$, with the remaining quantities defined below.

It is seldom that students encounter all three formulations during their studies, although they provide an instructive example of how identical underlying physics can be expressed in very different mathematical and conceptual forms. The Arrhenius equation is universally introduced in secondary school and undergraduate chemistry education, the Eyring formulation of transition state theory is standard in undergraduate physical chemistry curricula,\cite{Atkins,McQuarrieSimon,Laidler} whereas Chandler’s reactive flux perspective is typically encountered only at the graduate level or in specialized courses on statistical mechanics and molecular simulation.\cite{ChandlerText,FrenkelSmit2023}
One of the few modern textbooks that explicitly treats both perspectives is Peters.\cite{Peters2017}

The Arrhenius equation reveals the temperature dependence most straightforwardly, especially when the activation energy $E_a$ and the prefactor $A$ are treated as constants. This implies that a plot of $\ln k$ versus $1/T$ should be linear, from which $E_a$ and $A$ can be obtained from the slope and intercept, respectively. 
However, it is generally acknowledged that both $A$ and $E_a$ may not be strictly constant and can exhibit a \emph{weak} temperature dependence, the meaning of which will be discussed later.

The other two expressions do not directly expose the temperature dependence of the rate, since this requires specifying how the Gibbs free energy varies with temperature; nevertheless, they provide additional insight into how the rate constant depends on both system-specific and universal parameters. 
As we will show, these latter two expressions are formally equivalent. Their seemingly different appearance largely originates from the somewhat opaque $\Delta G^\ddagger$ term in the Eyring equation, into which several contributions appearing explicitly in Chandler’s reactive flux expression have effectively been absorbed.

The specific form of the Eyring equation has also led to several common misconceptions: (i) that it is inherently quantum mechanical because the Planck constant appears in the prefactor, (ii) that $\Delta G^\ddagger$ can be interpreted as the logarithm of a ratio of probabilities (or probability densities) in the same way as ordinary free energy differences, and (iii) that the temperature dependence can generally be written as $\Delta G^\ddagger(T)=\Delta H^\ddagger-T\Delta S^\ddagger$, with both $\Delta H^\ddagger$ and $\Delta S^\ddagger$ assumed temperature independent, or at most only weakly temperature dependent.

The latter misconception has contributed to the view that a plot of $\ln(k/T)$ versus $1/T$, allowing $\Delta H^\ddagger$ and $\Delta S^\ddagger$ to be obtained from the slope and intercept, is more fundamental than the Arrhenius representation.
The usual argument is that the Eyring equation follows from theoretical principles, whereas the Arrhenius equation is often regarded as merely phenomenological. In the following, however, I discuss the caveats underlying this reasoning and show how the equations can be reformulated such that activation enthalpies and entropies follow naturally from a linear Arrhenius representation. I further argue that these quantities are physically more meaningful than the corresponding values extracted from an Eyring plot.

\section{Theoretical Analysis}

\subsection{General remarks}
We will make several simplifying system assumptions for pedagogical reasons, leading to more transparent and intuitive expressions rather than fully generic but technically more cumbersome forms. These simplifications do not affect the validity of the conclusions, which also hold in more general formulations. The assumptions are listed below.

\emph{Order of the reaction.} All rate constants $k$ in Eq.~\ref{eq:3k} have units of s$^{-1}$, corresponding to an effective first-order decay from a metastable state $A$ in configuration space to a product state $B$. This interpretation also applies to reactions involving multiple molecular species by treating the molecular system as a whole, with higher-order rate constants recovered through straightforward concentration conversions.

\emph{Thermodynamic potential.}
The equations are expressed in terms of the Gibbs free energy $G$, which is customary in chemistry and has the practical advantage of avoiding the common ambiguity in the notation of the Helmholtz free energy, variously denoted by $F$ or $A$. To simplify matters, however, we assume a system in which no $pV$ work is performed, so that internal energy and enthalpy changes coincide and Gibbs and Helmholtz free energy differences become identical.

\emph{The reaction coordinate.} We assume the existence of a reaction coordinate $q$ for which the main TST assumptions hold, in particular that it provides a valid dividing surface between reactant and product states at $q=q^\ddagger$ with no recrossings. We further assume that the coordinate transformation from Cartesian coordinates to $q$ and the remaining orthogonal coordinates $\mathbf{q}^\perp$ has unit Jacobian, and that $q$ can be associated with a simple effective mass $m_q$ with units of kg. One may, for example, think of the $x$-coordinate of a particle diffusing in a nanoporous material with cubic symmetry, where transitions occur between well-separated cages. While this picture is most transparent for linear reaction coordinates, more general nonlinear coordinates lead to more involved mathematical forms. The corresponding effective mass is then defined through the coordinate transformation and need not correspond directly to a particle mass or retain simple physical units.\cite{Berne1988, Carter1989}

\subsection{The Arrhenius equation}

The validity of the Arrhenius equation is not necessarily restricted to TST, since deviations from ideal TST behavior can be incorporated through a transmission coefficient $\kappa$, which can effectively be absorbed into the prefactor $A$. Consequently, even when TST does not strictly apply, a linear Arrhenius plot may still be observed, leading to
\begin{align}
\ln(\tau k) = \ln(\tau A) - \frac{E_a}{k_B } \left(\frac{1}{T}\right)
\label{eq:linArr}
\end{align}
where $\tau$ is an arbitrary constant with units of seconds introduced to make the argument of the logarithms dimensionless. Eq.~\ref{eq:linArr} shows that $\ln \tau k$ versus $1/T$ is linear with slope $-E_a/k_B$ and intercept $\ln(\tau A)$.

However, it is  recognized that both $A$ and $E_a$ may exhibit a weak temperature dependence. 
We will say that a function $f(T)$ is weakly temperature dependent if both $f(T)$ and its inverse vary sublinearly with temperature, in the sense that $f(T)/T \to 0$ and $1/(T f(T)) \to 0$ for $T \to \infty$.
As an illustration, if $E_a(T)=b+aT$, the linear term can be absorbed into the prefactor, yielding an effective barrier $b$ and a modified prefactor $A\exp(-a/k_B)$. 
However, since this type of decomposition is exactly what is produced by the Arrhenius fitting procedure, $E_a(T)$ is essentially sublinear by construction, at least when the Arrhenius plot is approximately linear.

\subsection{Chandler's reactive flux}

In the reactive flux framework, the free energy profile is directly related to the logarithm of the probability density:
\begin{eqnarray}
G(q) = -k_B T \ln \left( \frac{\rho(q)}{\rho_{\rm ref}} \right)
\label{eq:Gq}
\end{eqnarray}
Here, $\rho_{\rm ref}$ defines the “zero offset” of the free energy and can be chosen arbitrarily, as only free energy differences have physical meaning.

The probability density is given by
\begin{eqnarray}
\rho(q') &=& \left\langle \delta(q-q') \right\rangle
= \frac{
\int \delta(q-q') e^{-\beta V({\bf q})} d{\bf q}
}{
\int e^{-\beta V({\bf q})} d{\bf q}
} \nonumber \\
&=&
\frac{
\int \delta(q-q') e^{-\beta V(q, {\bf q^\perp})} dq\, d{\bf q^\perp}
}{
\int e^{-\beta V({\bf q})} d{\bf q}
} \nonumber \\
&=&
\frac{
\int e^{-\beta V(q', {\bf q^\perp})} d{\bf q^\perp}
}{
\int e^{-\beta V({\bf q})} d{\bf q}
}
\label{eq:rho}
\end{eqnarray}
where we split the multidimensional configuration vector ${\bf q}$ into a scalar reaction coordinate $q$ and the remaining orthogonal coordinates ${\bf q^\perp}$. Here, $\delta(\cdot)$ is the Dirac delta function and $V({\bf q})$ is the potential energy surface.
Hence, the rate expression of Eq.~\ref{eq:3k} can be rewritten as
\begin{eqnarray}
k =
\sqrt{\frac{k_B T}{2 \pi m_q}}
\left(
\frac{
\int e^{-\beta V(q^\ddagger, {\bf q^\perp})} d{\bf q^\perp}
}{
\int_A e^{-\beta V({\bf q})} d{\bf q}
}
\label{eq:chandler}
\right)
\end{eqnarray}
where the subscript $A$ in the integral denotes integration over configuration space ${\bf q}=(q,{\bf q^\perp})$ restricted to $q \leq q^\ddagger$.
A derivation of this expression is provided in the Supportive Information (SI).

 \subsection{Eyring equation}

Historically, the Eyring equation was derived by interpreting the transition state as a loosely bound activated complex possessing a very low vibrational frequency $\nu^\ddagger$. We now understand, however, that a true transition state corresponds to a saddle point on the potential energy surface with negative curvature along the reaction coordinate and therefore an imaginary frequency. 
Interestingly, simplified textbook presentations of this picture are still common in physical chemistry textbooks. In Atkins,\cite{Atkins} an explicit caveat that this interpretation is an oversimplification was present in several earlier editions but is absent from the two most recent editions.\footnote{The caveat was present in the 6th, 9th, and 10th editions examined by the author, but not in the 11th and 12th editions. Its removal is disconcerting because students are no longer explicitly warned that the characteristic dip in the free-energy profile is an artifact of the representation rather than a genuine feature of the transition state.}
In the original derivation by Eyring,\cite{Eyring1935} the contribution of the hypothetical soft vibrational mode is treated by assuming $\beta h \nu^\ddagger \ll 1$, motivated by the assumption that $\nu^\ddagger$ is very small, and retaining only the leading term in the corresponding Taylor expansion. This leaves a linear dependence on $\nu^\ddagger$ that subsequently cancels. 
In effect, this corresponds to the classical limit $h \to 0$, i.e. retaining only the leading-order contribution in $h$, since the same approximation follows 
when $h$ is small compared to the characteristic action scale of the system, equivalently when $\beta h \nu^\ddagger \ll 1$.
Despite the somewhat questionable intermediate picture of a loosely bound vibration, the resulting rate expression is correct and can be derived more rigorously in several alternative ways,\cite{Laidler} including purely classical statistical mechanical approaches.\cite{Mahan}

Here, we derive the Eyring equation within a classical statistical mechanical framework by showing its equivalence to Eq.~\ref{eq:chandler}, whose derivation is provided in the SI. To this end, we start from the standard partition function $Q$, in which the Planck constant appears as a normalization factor,
\begin{eqnarray}
Q=\frac{1}{h^{3N}} \int e^{-\beta H({\bf q}, {\bf p})} \, d{\bf q}\, d{\bf p}
\end{eqnarray}
where $H$ is the classical Hamiltonian and ${\bf p}$ denotes the momenta. For simplicity, we omit the factorial term $1/N!$ (or a product of factorials for systems containing multiple atom types). This corresponds to adopting the labeled-microstate convention, in which permutations of identical particles are counted as distinct microstates because each particle is assigned a label.

The normalization factor $h^{3N}$ is conventionally introduced\cite{Hill_Q} such that the classical partition function approaches the corresponding quantum-mechanical partition sum in the limit $h \rightarrow 0$ (see SI Eqs.~S9--S11, where this is shown for a harmonic oscillator). It is important to recognize, however, that this normalization is not a physical necessity but a conventional choice. For any experimentally measurable observable, only relative Boltzmann weights are relevant. The Boltzmann factor determines that the probability of a microstate, whether a discrete quantum state or a phase-space point in the classical continuum, is proportional to $e^{-\beta H}$. The partition function merely provides the normalization of this probability distribution and therefore has no direct physical meaning through its absolute value alone.

For a system of particles with identical mass, we can integrate out the momenta (via Eq.~S1),
\begin{align}
Q
&= \frac{1}{h^{3N}} \int e^{-\beta V({\bf q})}\, d{\bf q}
\int e^{-\beta {\bf p}^2/2m}\, d{\bf p}  \\
&= \int e^{-\beta V({\bf q})}\, d{\bf q}
\left(\frac{\sqrt{2\pi m k_B T}}{h}\right)^{3N} \nonumber \\
&= \int e^{-\beta V({\bf q})}\, d{\bf q}
\left(\frac{1}{\Lambda}\right)^{3N},
\quad
\Lambda = \sqrt{\frac{h^2}{2\pi m k_B T}}.\nonumber
\end{align}
Here, $\Lambda$ is the thermal de Broglie wavelength. For a system with different particle masses, the same procedure applies independently to each particle, resulting in a product of mass-dependent thermal wavelengths, $\prod_i \Lambda_i^{-3}$, where $\Lambda_i = \sqrt{h^2/(2\pi m_i k_B T)}$. To keep the notation compact, one may formally write $\Lambda$ as a geometric mean of the individual thermal wavelengths. However, when considering a specific degree of freedom, the corresponding mass associated with that degree of freedom must be used in the definition of $\Lambda_i$.

 We can now introduce the partition function of state $A$ by restricting the configuration integral to the reactant region,
\begin{align}
Q_A
= \int_A e^{-\beta V({\bf q})}\, d{\bf q}
\left(\frac{1}{\Lambda}\right)^{3N}.
\end{align}

We next define the transition-state partition function by removing the phase-space degrees of freedom associated with the reaction coordinate, effectively eliminating one $dq\,dp/h$ contribution and fixing $(q,p)$ to $(q^\ddagger,0)$. This yields
\begin{align}
Q_{\ddagger}
&= \int e^{-\beta V(q^\ddagger,{\bf q^\perp})}\, d{\bf q^\perp}
\left(\frac{1}{\Lambda}\right)^{3N-1}.
\end{align}

Here, we allowed ourselves a slightly informal notation to keep the expressions compact.
As stated above, in the presence of mass polydispersity, $\Lambda$ depends on the mass associated with each degree of freedom. Removing the reaction coordinate therefore removes one specific thermal wavelength from the product. Consequently, the ratio
$\left(\frac{1}{\Lambda}\right)^{3N-1}/\left(\frac{1}{\Lambda}\right)^{3N}$
corresponds to the thermal wavelength $\Lambda_q$ associated with the reaction coordinate $q$, defined through the effective mass $m_q$ of that degree of freedom.

Then, by defining the absolute free energies of the reactant state and the transition state via $Q_A$ and $Q_{\ddagger}$, with $G_X = -k_B T \ln Q_X$, we obtain
\begin{align}
\Delta G^\ddagger
&= G_{\ddagger} - G_A
= -k_B T \ln\!\left(\frac{Q_{\ddagger}}{Q_A}\right)  \label{eq:DGQQ}\\
&= -k_B T \ln\!\left(
\frac{
\Lambda_q \int e^{-\beta V(q^\ddagger, {\bf q^\perp})}\, d{\bf q^\perp}
}{
\int_A e^{-\beta V({\bf q})}\, d{\bf q}
}
\right).\nonumber
\end{align}
Hence, using $\Lambda_q = \sqrt{h^2/(2\pi m_q k_B T)}$:
\begin{align}
\frac{k_B T}{h} e^{-\beta \Delta G^\ddagger}
=
\sqrt{\frac{k_B T}{2\pi m_q}}
\left(
\frac{
\int e^{-\beta V(q^\ddagger, {\bf q^\perp})}\, d{\bf q^\perp}
}{
\int_A e^{-\beta V({\bf q})}\, d{\bf q}
}
\right),\nonumber
\end{align}
which is identical to Eq.~\ref{eq:chandler}. Thus, Chandler’s expression and Eyring’s equation are equivalent when the microscopic definition of $\Delta G^\ddagger$ is fully resolved.

Up to this point, we have shown that the Eyring expression is mathematically identical to Chandler’s reactive-flux expression. Since the latter was derived rigorously in the SI, this also establishes the validity of the Eyring equation itself. However, the situation becomes less clear once additional assumptions about $\Delta G^\ddagger$ are introduced, as is commonly done in the construction and interpretation of Eyring plots.

\subsection{Harmonic Approximation}

The Eyring plot relies on the assumption that, when $\Delta G^\ddagger$ is decomposed into the enthalpy and entropy of activation according to
\begin{align}
\Delta G^\ddagger &= \Delta H^\ddagger - T \Delta S^\ddagger,\nonumber \\
\Rightarrow\quad 
k &=
\frac{k_B T}{h} \,e^{\Delta S^\ddagger/k_B} \, e^{-\beta \Delta H^\ddagger},
\label{eq:DGDHDS}
\end{align}
both $\Delta H^\ddagger$ and $\Delta S^\ddagger$ are temperature independent, or at most only weakly temperature dependent.
Under this assumption, plotting $\ln(k/T)$ against $1/T$ yields a linear relation:
\begin{align}
    \ln \left( \frac{\gamma k}{T} \right)
    =
    \ln \left( \frac{\gamma k_B}{h} \right)
    + \frac{\Delta S^\ddagger}{k_B}
    - \frac{\Delta H^\ddagger}{k_B}\left(\frac{1}{T}\right) \nonumber
\end{align}
where $\gamma$ is an arbitrary constant with units  s$\cdot$K.

In practice, however, the temperature dependence of the prefactor, when written as $\propto T^n$, is often too weak to be resolved within the usual experimental uncertainty for values of $n$ between $0$ and $1$.\cite{Petrou2012ChemGeo} Consequently, data that appear linear in an Arrhenius plot will generally also appear linear in an Eyring plot within the same experimental uncertainty.\cite{Petrou2012ChemGeo}

\begin{figure}[ht]
    \centering
    \includegraphics[width=.99\linewidth]{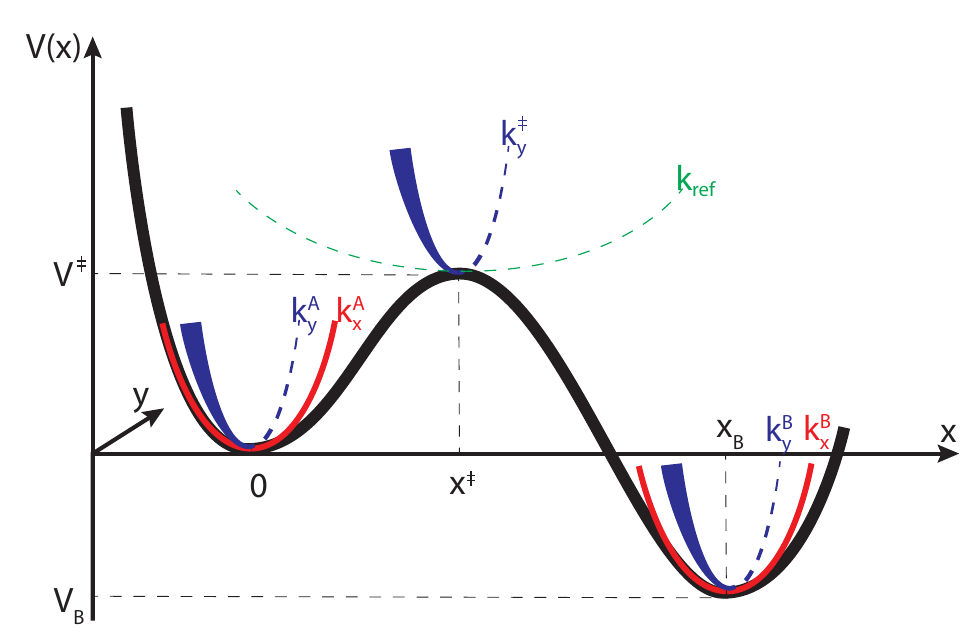}
    \caption{
    Two-dimensional potential energy surface.
    Curvatures along the reaction coordinate $x$ in the reactant and product wells are indicated in red.
    Curvatures along the orthogonal coordinate $y$ at the reactant, transition, and product states are shown in blue.
    The hypothetical reference curvature $k_{\mathrm{ref}}$ discussed in the text is indicated by the dashed green line.
   }
    \label{fig:2Dharmonic}
\end{figure}

Therefore, because it is experimentally difficult to determine unambiguously whether an Arrhenius or an Eyring representation provides the more appropriate linear relationship, it is necessary to analyze theoretical models for which the rate constant can be computed analytically. To this end, we consider a two-dimensional potential energy surface (Fig.~\ref{fig:2Dharmonic}), 
where the reactant ($A$), transition ($\ddagger$), and product ($B$) regions can, for $y \approx 0$, be approximated by harmonic quadratic expansions around their stationary points:
\begin{align}
V \approx
\left\{
\begin{array}{ll}
\frac{1}{2} k_x^A x^2 + \frac{1}{2} k_y^A y^2, 
& x \approx 0, \\[4pt]
V^\ddagger
+ \frac{1}{2} k_y^{\ddagger} y^2, 
& x \approx x^\ddagger, \\[4pt]
V_B + \frac{1}{2} k_x^B (x-x_B)^2 + \frac{1}{2} k_y^B y^2, 
& x \approx x_B
\end{array}
\right.\nonumber
\end{align}
The resulting rate constant can be obtained within classical harmonic 
transition state theory (HTST),\cite{Peters2017} with a derivation given in the SI, yielding
\begin{align}
k=
\frac{\nu_x^A \nu_y^A}{\nu_y^\ddagger}
\exp(-\beta V^{\ddagger}),
\end{align}
which can be identified with Arrhenius parameters as
$E_a = V^\ddagger$ and
$A = \left(\nu_x^A \nu_y^A / \nu_y^\ddagger\right)$.
Substituting this expression into Eq.~\ref{eq:3k} gives
\begin{align}
\Delta G^\ddagger
=
V^\ddagger
-
k_B T
\ln\left(
\frac{h \nu_x^A \nu_y^A}{k_B T \nu_y^\ddagger}
\right).
\label{eq:DG++HTST}
\end{align}
and a seemingly natural decomposition follows 
\begin{align}
\Delta \tilde{H}^\ddagger = V^\ddagger, 
\quad
\Delta \tilde{S}^\ddagger =
k_B
\ln\left(
\frac{h \nu_x^A \nu_y^A}{k_B T \nu_y^\ddagger}
\right),
\label{eq:wrong_split}
\end{align}
such that the enthalpy of activation equals the barrier height and, equivalently, the Arrhenius activation energy $E_a$. This choice therefore satisfies the requirement of weak temperature dependence; indeed, $\Delta \tilde{H}^\ddagger$ is constant. This decomposition is, however, thermodynamically inconsistent, as discussed below.

If we inspect $\Delta \tilde{S}^\ddagger$, we observe that the frequencies along the orthogonal degree of freedom $y$ contribute to the entropy term whenever they differ between reactant and transition state, while leaving the enthalpy term unaffected. This can be understood intuitively: a small frequency $\nu_\alpha^X$ corresponds to a low curvature along direction $\alpha$ at location $X$, implying a relatively flat potential-energy landscape and thus a broad region in configuration space where the system can move without significant energetic penalty.
Although the temperature dependence of $\Delta \tilde{S}^\ddagger$ may appear weak, the factor $\exp(\Delta \tilde{S}^\ddagger/k_B)$ is not weakly temperature dependent; in fact, it scales as $\propto 1/T$, thereby compensating the apparent linear temperature dependence in Eq.~\ref{eq:DGDHDS}.

However, Eq.~\ref{eq:wrong_split} does not satisfy the Gibbs--Helmholtz relation, which can be written as $d (\beta \Delta G^\ddagger)/d\beta = \Delta H^\ddagger$. Enforcing this relation yields
\begin{align}
\Delta H^\ddagger(T) &= V^\ddagger - k_B T, \nonumber \\
\Delta S^\ddagger(T) &= k_B \left[ \ln \left( 
\frac{h \nu_x^A \nu_y^A}{k_B T \nu_y^\ddagger} 
 \right) - 1 \right],
\label{eq:right_split}
\end{align}
with $\Delta H^\ddagger$ 
exhibiting a contribution that is linear in $T$.

Eq.~\ref{eq:right_split} also provides insight into the controversial phenomenon of enthalpy--entropy compensation.\cite{CornishBowden2002} It has been argued that the observed correlations between $\Delta H^\ddagger$ and $\Delta S^\ddagger$ mainly arise from correlated fitting errors, since both quantities are extracted from the same Eyring plot, suggesting that the effect may be largely artefactual.\cite{CornishBowden2002,Petersen1961SolventEI, McBane1998} However, as shown in the SI, enthalpy--entropy compensation also appears in the present exactly solvable model, i.e., in the absence of experimental uncertainty. Nevertheless, even in this case the phenomenon originates from the procedure itself, namely fitting a straight line to an inherently slightly nonlinear curve, leading to a correlation that is mathematically real but not physically meaningful. Therefore, in the next section, we examine the physical interpretation of $\Delta G^\ddagger$ in more detail.

\subsection{Interpretation of $\Delta G^\ddagger$} 

A natural way to think about free energy is as a measure of statistical likelihood. More specifically, the absolute free energy can be viewed as the logarithm of an equilibrium statistical weight (i.e., probability or probability density), so that free energy differences correspond to logarithms of ratios of these weights. In this picture, more likely states correspond to lower free energies, whereas less likely states correspond to higher free energies. This interpretation leads to ratios of partition functions over configuration space, unless the reaction coordinate $q$ depends on momenta, which is generally not the case.

For instance, the reaction free energy $\Delta G_{AB}$ relates to the probabilities of the product $B$ and reactant $A$ states:
\begin{align}
\Delta G_{AB} 
&= G_B-G_A=-k_B T \ln \left( \frac{P_B}{P_A} \right)
\nonumber \\
&= -k_B T \ln
\left(
\frac{
\int_B e^{-\beta V({\bf q})}\, d{\bf q}
}{
\int_A e^{-\beta V({\bf q})}\, d{\bf q}
}
\right),\label{eq:DGPAPB}
\end{align}
where the second line expresses the ratio in terms of configurational partition functions restricted to the reactant region $A$ and the product region $B$.

 Likewise, the minimum–maximum free energy difference obtained from a free energy profile (FEP) along a reaction coordinate is written as
\begin{align}
\Delta G^\ddagger_{\rm FEP}
&=
-k_B T \ln \left(
\frac{\rho(q^\ddagger)}{\rho(q_A)}
\right),\nonumber \\
&=-k_B T \frac{
\int e^{-\beta V(q^\ddagger, {\bf q^\perp})} d{\bf q^\perp}
}{
\int e^{-\beta V(q_A, {\bf q^\perp})} d{\bf q^\perp}
}\label{eq:DGrho}
\end{align}
where $q_A$ denotes the position of the minimum of the reactant well. Here, $\rho(q)$ is a probability density rather than a probability and therefore carries units inverse to those of the reaction coordinate $q$. The units cancel in the ratio $\rho(q^\ddagger)/\rho(q_A)$, making the free energy difference well defined and physically interpretable.

In contrast, $\Delta G^\ddagger$ in the Eyring framework attempts to define a free energy difference between fundamentally different objects: a reactant state with finite probability $P_A$, and a transition state defined as a hypersurface with probability density $\rho(q^\ddagger)$. A direct ratio $\rho(q^\ddagger)/P_A$ is therefore ill-defined, as it either carries units or corresponds to comparing a vanishing probability of a hypersurface with a finite one. More precisely, Eyring’s transition state formulation introduces an additional restriction on the momentum along the reaction coordinate, yielding a phase-space density at the dividing surface, $\rho(q=q^\ddagger, p=0)$. Taking the ratio with $P_A$ leads to
\begin{align}
\Delta G^\ddagger=
-k_B T \ln \left( h \cdot
\frac{\rho(q=q^\ddagger, p=0)}{P_A}
\right),\label{eq:DGrhoqp}
\end{align}
where the factor $h$ renders the argument of the logarithm dimensionless. However, as argued above, the introduction of this constant is ultimately a matter of convention, arising from the normalization of the semiclassical partition function $Q$, rather than a physical necessity, since any quantity with dimensions of action (J\,s or equivalently kg\,m$^2$s$^{-1}$) could equally well be used.
Such a choice would simply 
change the definition of $\Delta G^\ddagger$ and
shift the corresponding factor into the prefactor of the Eyring rate expression.

This highlights the conceptual clarity gained by interpreting free energy differences directly in terms of their physically meaningful basis, namely as logarithms of probabilities or probability densities. At the more abstract level of partition function ratios, especially when these are rendered dimensionless by construction, the subtle issue of comparing quantities with different dimensionality becomes largely hidden. As a result, expressions such as $Q_{\ddagger}/Q_A$ in Eq.~\ref{eq:DGQQ} tend to obscure both the underlying dimensionality mismatch and the fact that $\Delta G^\ddagger$ is not as uniquely or unambiguously defined as the free energy differences in Eqs.~\ref{eq:DGPAPB}--\ref{eq:DGrho}.

Indeed, $\Delta G^\ddagger$ does not behave in all respects in the same way as the other free energies. This becomes evident when evaluating Eqs.~\ref{eq:DGPAPB}--\ref{eq:DGrho} within the harmonic approximation,
\begin{align}
    \Delta G_{AB}&=V_B - 
    k_B T \ln \left( 
    \frac{\nu_x^A \nu_y^A}{
    \nu_x^B \nu_y^B
    }\right), \nonumber\\
    \Delta G_{\rm FEP}^\ddagger&=
     V^\ddagger - k_B T \ln \left( \frac{\nu^A_y}{\nu^{\ddagger}_y} \right),
     \label{eq:GABGFEP}
\end{align}
which can be derived in a manner fully analogous to the previous results. These expressions can be decomposed into enthalpic and entropic contributions,
\begin{align}
\Delta H_{AB} &= V_B, \quad
\Delta S_{AB} = 
 k_B \ln \left( 
    \frac{\nu_x^A \nu_y^A}{
    \nu_x^B \nu_y^B
    }\right), \nonumber\\
\Delta H_{\mathrm{FEP}}^\ddagger &= V^{\ddagger}, \quad
\Delta S_{\mathrm{FEP}} = k_B \ln \left( \frac{\nu^A_y}{\nu^{\ddagger}_y} \right).
\end{align}
which are temperature independent, unlike $\Delta H^\ddagger$ and
$\Delta S^\ddagger$ 
(Eq.~\ref{eq:right_split}).

The commonly used assumption that $\Delta H^\ddagger$ is temperature independent, or at most weakly temperature dependent, is based on the assumption of similar molar heat capacities in reactant and transition states. This assumption is reasonable when the states being compared have the same dimensionality, as in $H_{AB}$ and $\Delta H_{\rm FEP}^\ddagger$. However, for $\Delta H^\ddagger$, the reactant well contains additional spatial and momentum degrees of freedom relative to the transition-state representation. These give rise to quadratic energy contributions, each of order $\tfrac{1}{2}k_B T$, which are not canceled as in the other cases. This difference limits the validity of treating $\Delta H^\ddagger$ as temperature independent and, consequently, weakens the physical basis of Eyring plots (i.e., linear fits of $\ln (k/T)$ versus $1/T$).

Comparing Eqs.~\ref{eq:GABGFEP} and \ref{eq:DG++HTST}, we obtain
\begin{equation}
\Delta G^\ddagger
= \Delta G_{\mathrm{FEP}} - k_B T \ln \left( \frac{\nu^R_x h}{k_B T} \right).
\end{equation}
This expression shows that the common pragmatic approach of extracting a minimum--maximum free energy difference from a FEP and directly inserting it into Eyring’s equation is formally incorrect.\cite{Bocus2026}
To estimate the magnitude of the bias, consider a typical O--H vibrational mode with $\tilde{\nu} \approx 3400~\mathrm{cm}^{-1}$, corresponding to $\nu \approx 10^{14}~\mathrm{s}^{-1}$. At $T = 300~\mathrm{K}$, this gives $\nu h / k_B T \approx 16$. Hence, replacing $\Delta G^\ddagger$ by $\Delta G_{\mathrm{FEP}}$ would lead to an underestimation of the rate by a factor of approximately 16.
Ironically, this approximation may in some cases appear closer to experimental results when a low-quality reaction coordinate is used, since such coordinates typically yield a $\Delta G_{\mathrm{FEP}}$ that is significantly lower than the true kinetic barrier. This can lead to a fortuitous cancellation of errors.

\subsection{Alternative dimensionality fix}

We have now exposed several shortcomings of the Eyring plot. Although the Eyring equation itself is correct, the Eyring plot will in general not be linear, and the definition of $\Delta G^\ddagger$ is based on a conventional choice introduced to account for the mismatch in dimensionality between the transition state and the reactant state. We have also seen that this leads to undesirable properties of the likewise convention-dependent quantities $\Delta H^\ddagger$ and $\Delta S^\ddagger$, in particular their notable temperature dependence.

At the same time, there are important advantages of the Eyring formulation compared to the formulations based on the Arrhenius equation and Chandler's reactive flux, Eq.~\ref{eq:3k}, especially from the perspective of experimental researchers. The Arrhenius prefactor $A$ does not have the same conceptual appeal as $\Delta S^\ddagger$, while Chandler’s expression, although in principle more intuitive and free from convention-defined quantities, contains so many variables that it becomes difficult to apply in practice for the interpretation of experiments. For instance, it is based not on a single free energy difference but on a full free energy profile. In addition, it is difficult to determine the mass $m_q$ associated with the ideal reaction coordinate, a coordinate that is generally not known beforehand and is already difficult to obtain in molecular computer simulations, let alone in experiments. While HTST could in principle be combined with accurate vibrational spectroscopy measurements, in many cases this will be difficult, and the harmonic approximation tends to break down at high temperatures.

The appeal of the Eyring plot is precisely that it allows experimental researchers to extract activation enthalpies and entropies using relatively simple means, namely by measuring the rate as a function of temperature. Furthermore, even if $\Delta S^\ddagger_{\mathrm{FEP}}$ can be determined from molecular simulations, it does not include the entropic contribution along the reaction coordinate itself, as $\Delta S^\ddagger$ does. The question I would therefore like to raise is whether it is possible to redefine the convention for $\Delta G^\ddagger$ in such a way that these practical advantages are retained while avoiding the drawbacks associated with the current convention.

If we compare $\Delta G^\ddagger$ in Eq.~\ref{eq:DG++HTST} with $G_{AB}$ in Eq.~\ref{eq:GABGFEP}, we find that $\Delta G^\ddagger$ behaves as if it were a convention-free free energy difference such as $G_{AB}$, but with a transition state that effectively possesses a temperature-dependent vibrational mode given by $\nu_x^{\ddagger} = k_B T / h$. In this picture, the unstable mode along $x$, characterized by a negative curvature and an imaginary frequency, is replaced by a stable mode with positive curvature and an effective vibrational frequency that is real, positive, and temperature dependent. This curvature must scale with mass in order to yield a mass-independent prefactor in the Eyring equation, but at the same time introduces an undesirable temperature dependence.

We therefore propose to explicitly acknowledge that a free energy of activation cannot be defined without a convention, and instead choose this convention such that the resulting expression has the desired properties. Specifically, we introduce a hypothetical transition state that matches the dimensionality of the reactant state by adding a harmonic 
curvature
along $q$, centered at $q^\ddagger$, with a prescribed reference frequency $\nu_{\mathrm{ref}}$. The corresponding curvature is hence $k_{\mathrm{ref}} = m_q (2 \pi \nu_{\mathrm{ref}})^2$, and the transition-state potential becomes
$V^\ddagger_{\mathrm{ref}}(q, \mathbf{q}^\perp) = V(q^\ddagger, \mathbf{q}^\perp) + \frac{1}{2} k_{\mathrm{ref}} (q - q^\ddagger)^2$.
In this construction, motion along the added mode is assumed to be classical. This choice fixes the convention and allows us to rewrite Eq.~\ref{eq:chandler} as
\begin{align}
k
&=
\sqrt{\frac{k_B T}{2 \pi m_q}}
\\
&\times
\left(
\frac{
\int e^{-\beta V_{}(q^\ddagger,\mathbf{q^\perp})} \, d\mathbf{q^\perp} \int_{-\infty}^{\infty}
e^{-\beta \frac{1}{2} k_{\rm ref} (q^\ddagger - q)^2} \, dq
}{
\int_A e^{-\beta V(\mathbf{q})} \, d\mathbf{q}
\int_{-\infty}^{\infty}
e^{-\beta \frac{1}{2} k_{\rm ref} (q^\ddagger - q)^2} \, dq
}
\right)
\nonumber
\\
&=
\sqrt{\frac{k_B T}{2 \pi m_q}}
\left(
\frac{
\int e^{-\beta V_{\rm ref}^\ddagger(\mathbf{q})} \, d\mathbf{q}
}{
\int_A e^{-\beta V(\mathbf{q})} \, d\mathbf{q}
}
\right)
\left(
\frac{1}{\sqrt{\frac{2 \pi k_B T}{k_{\rm ref}}}}
\right) \nonumber \\
&=
\nu_{\rm ref}
\left(
\frac{
\int e^{-\beta V_{\rm ref}^\ddagger(\mathbf{q})} \, d\mathbf{q}
}{
\int_A e^{-\beta V(\mathbf{q})} \, d\mathbf{q}
}
\right)
=
\nu_{\rm ref} \, e^{-\beta \Delta G^\ddagger_{\rm ref}}.\nonumber
\end{align}
where Eq.~S1 was used to evaluate the Gaussian integral.

Several choices for $\nu_{\rm ref}$ are possible. One could, for example, use a well-defined molecular vibrational frequency (such as an OH stretching mode), simply set $\nu_{\rm ref} = 1~\mathrm{s}^{-1}$, or adopt a value inspired by the Eyring prefactor, for instance $\nu_{\rm ref} = k_B T_{\rm ref}/h$, where $T_{\rm ref}$ is a fixed reference temperature, such as room temperature.

Now, the key advantage of this formulation is that both the reactant state and the transition state are defined on spaces of equal dimensionality. As a consequence, decomposing $\Delta G^\ddagger_{\rm ref}$ as $\Delta H^\ddagger_{\rm ref} - T \Delta S^\ddagger_{\rm ref}$ renders the assumption of temperature-independent $\Delta H^\ddagger_{\rm ref}$ and $\Delta S^\ddagger_{\rm ref}$ significantly more justified and  it is even exact within the harmonic approximation.

As a result, an Arrhenius-type linear plot is expected following
\begin{eqnarray}
\ln (k/v_{\rm ref}) &=&
 \frac{\Delta S^\ddagger_{\rm ref}}{k_B}
-
\frac{\Delta H^\ddagger_{\rm ref}}{k_B}
\left(\frac{1}{T}\right) 
\end{eqnarray}

The definitions of $\Delta G_{\rm ref}^\ddagger$,
$\Delta H_{\rm ref}^\ddagger$, and
$\Delta S_{\rm ref}^\ddagger$
can be expressed directly in terms of configurational partition integrals. For
$\Delta G_{\rm ref}^\ddagger$, we obtain
\begin{align}
    \Delta G_{\rm ref}^\ddagger
    &=-k_B T \ln
    \left(
\frac{
\int e^{-\beta V_{\rm ref}^\ddagger(\mathbf{q})} \, d\mathbf{q}
}{
\int_A e^{-\beta V(\mathbf{q})} \, d\mathbf{q}
}
\right) \\
&=-k_B T \ln
    \left(
\frac{
\int e^{-\beta V(q^\ddagger,\mathbf{q^\perp})} \, d\mathbf{q^\perp} \sqrt{
\frac{2 \pi k_B T}{k_{\rm ref}}
}
}{
\int_A e^{-\beta V(\mathbf{q})} \, d\mathbf{q}
}
\right)\nonumber \\
&=-k_B T
  \ln  \left(
\frac{
\sqrt{ 
\frac{k_B T}{2 \pi m_q}  }
\int e^{-\beta V(q^\ddagger,\mathbf{q^\perp})} \, d\mathbf{q^\perp} 
}{\nu_{\rm ref}
\int_A e^{-\beta V(\mathbf{q})} \, d\mathbf{q}
}
\right) \nonumber
\end{align}
For $\Delta H_{\rm ref}^\ddagger$, we similarly find
\begin{align}
    \Delta H_{\rm ref}^\ddagger
    &=\! 
    \left(
\frac{
\int \! V_{\rm ref}^\ddagger(\mathbf{q}) e^{-\beta V_{\rm ref}^\ddagger(\mathbf{q})}  d\mathbf{q}
}{
\int \!  e^{-\beta V_{\rm ref}^\ddagger(\mathbf{q}) }  d\mathbf{q}
}
\right) 
\\ &
\! -\! 
\left(
\frac{
\int_A \! V(\mathbf{q}) e^{-\beta V(\mathbf{q})} d\mathbf{q}
}{
\int_A \! e^{-\beta V(\mathbf{q})}  d\mathbf{q}
}
\right)  \nonumber \\
&= \left(
\frac{
\int V(q^\ddagger,\mathbf{q^\perp}) e^{-\beta 
V(q^\ddagger,\mathbf{q^\perp})
} \, d\mathbf{q^\perp}
}{
\int e^{-\beta V(q^\ddagger,\mathbf{q^\perp})
} \, d\mathbf{q^\perp}
}
\right) 
\nonumber \\ &
-
\left(
\frac{
\int_A V(\mathbf{q}) e^{-\beta V(\mathbf{q})} \, d\mathbf{q}
}{
\int_A e^{-\beta V(\mathbf{q})} \, d\mathbf{q}
}
\right) +\frac{1}{2} k_B T
\label{eq:DHref_partfunc}
\end{align}
which is independent of the particular choice of $\nu_{\rm ref}$. Finally,
    \begin{align}
    &\Delta S^\ddagger_{\rm ref}=
    \frac{\Delta H^\ddagger_{\rm ref}-
    \Delta G^\ddagger_{\rm ref}}{T}
\end{align}

 In the harmonic approximation, one obtains
\begin{align}
   & \Delta H^\ddagger_{\rm ref} = V^\ddagger, \quad
    \Delta S^\ddagger_{\rm ref} = 
    k_B \ln \left( 
    \frac{\nu_x^A \, \nu_y^A}{
    \nu_{\rm ref} \, \nu_y^\ddagger 
    }\right)
\end{align}
The activation enthalpy $\Delta H^\ddagger_{\rm ref}$ is identical to $E_a$ in the Arrhenius equation and is strictly independent of the arbitrary reference frequency $\nu_{\rm ref}$. The subscript ``ref'' is retained only to distinguish this definition from the Eyring convention.

The additional term $\tfrac{1}{2} k_B T$ in Eq.~\ref{eq:DHref_partfunc} arises from the effective constraining curvature $k_{\rm ref}$, which repairs the dimensionality mismatch. It leads to exact cancellation of the temperature dependence in the harmonic limit and leaves only a weak temperature dependence when anharmonicities are present. In the Eyring formulation, this stabilizing contribution is effectively replaced by $-\tfrac{1}{2} k_B T$, which instead enhances the artificial temperature dependence associated with the dimensionality mismatch.

The quantities $\Delta G^\ddagger_{\rm ref}$ and $\Delta S^\ddagger_{\rm ref}$ depend on the chosen reference frequency $\nu_{\rm ref}$, and this choice must therefore be reported explicitly when quoting activation free energies and entropies. However, changing $\nu_{\rm ref}$ only introduces a trivial additive shift: $\Delta G^\ddagger_{\rm ref,2} = \Delta G^\ddagger_{\rm ref,1} + k_B T \ln\!\left(\frac{\nu_{\rm ref,2}}{\nu_{\rm ref,1}}\right)$, and $\Delta S^\ddagger_{\rm ref,2} = \Delta S^\ddagger_{\rm ref,1} + k_B \ln\!\left(\frac{\nu_{\rm ref,2}}{\nu_{\rm ref,1}}\right)$. Consequently, values reported in the literature using different reference frequencies can be converted straightforwardly and compared directly.

\section{Conclusions}

In this article, I have highlighted several points related to transition state theory (TST) that are important both from an educational perspective and for scientific research. In particular, I have presented three equations, namely the Arrhenius equation, the Eyring equation, and Chandler's reactive flux expression, that can all be used to describe the temperature dependence of rate constants within TST, yet are rarely discussed together. As a result, few students are exposed to all three formulations, while even experienced researchers and educators are often familiar with the Arrhenius equation and only one of the other two expressions.

I believe this is a missed opportunity, as studying the Eyring equation and Chandler’s reactive flux expression together provides a valuable learning experience and a deeper understanding of how reaction rates depend on temperature and mass. This is particularly clear once it is shown, as done here, that the two expressions are formally equivalent despite initially suggesting different dependencies. Although the Eyring equation is often associated with quantum mechanics and Chandler’s expression with classical statistical mechanics, this distinction is largely historical. In fact, both expressions can be derived from classical mechanical principles. In both cases, the Gibbs free energy may include quantum mechanical contributions to varying extent through orthogonal degrees of freedom associated with rotational and vibrational modes, and this applies equally to both formulations.

A key element in understanding the equivalence of both expressions is the interpretation of the activation free energy $\Delta G^\ddagger$. Its definition depends on an implicit convention and cannot be derived purely from physical principles. $\Delta G^\ddagger$ compares statistical probabilities of the transition and reactant states, which are not directly comparable due to their different dimensionality, requiring a dimensionality correction with multiple valid conventions. In particular, the Planck constant in the Eyring prefactor is exactly compensated by its contribution to $\Delta G^\ddagger$, so the apparent quantum signature cancels in the final rate expression.

While the understandable confusion on this point can be viewed as harmless, more serious is the misunderstanding that the Eyring plot, $\ln(k/T)$ versus $1/T$, is inherently more fundamental and therefore preferred over the Arrhenius plot, $\ln k$ versus $1/T$. The repeated argument is that Arrhenius is merely phenomenological, while the Eyring equation is based on theoretical foundations. I find this reasoning somewhat amusing, as if it still matters that Arrhenius did not yet have a ready-made theoretical explanation in 1889, while ignoring our present understanding that allows us, from fundamental principles, to understand quite well why Arrhenius plots often appear linear. In contrast, even though the Eyring equation is sound, the assumption that an Eyring plot is linear is incorrect and reflects a poor understanding of the temperature dependence of $\Delta G^\ddagger$ (and the related $\Delta H^\ddagger$ and $\Delta S^\ddagger$). This originates from the aforementioned dimensionality mismatch, which leads to a non-standard temperature dependence compared to ordinary free energy differences such as reaction free energies.

In particular, the assumption leading to an expected linear Eyring plot is that $\Delta H^\ddagger$ and $\Delta S^\ddagger$ are temperature independent or only weakly dependent on temperature. Using harmonic TST, we show this is not valid: $\Delta H^\ddagger$ contains a term linear in $T$. Although $\Delta S^\ddagger$ might by itself be considered only weakly temperature dependent, this does not hold for the relevant contribution to the prefactor, $\exp(\Delta S^\ddagger/k_B)$. Instead, this term exhibits an inverse temperature dependence that cancels the apparent linearity in the Eyring expression. It is therefore more consistent to rely on the linearity of the Arrhenius plot and extract $\Delta H^\ddagger$ and $\Delta S^\ddagger$ from Arrhenius parameters, using $\Delta H^\ddagger = E_a - k_B T$ and $\Delta S^\ddagger = k_B \ln\left(A h/k_B T\right) - k_B$.

The argument that many experimental kinetic data sets appear linear in an Eyring plot is insufficient, since such data would typically also appear linear in an Arrhenius plot within experimental uncertainty. As long as experimental techniques cannot resolve such subtle deviations in logarithmic representations, one should rely on exactly solvable models such as harmonic TST, which unambiguously favors the Arrhenius representation. When anharmonicities are present, deviations are better described by a general expansion $A(T) = A_0 + A_1 T + A_2 T^2 + \ldots$, with no reason to assume dominance of a linear term unless dictated by a specific mechanism. Other temperature dependences may also arise; for example, hard-core models can exhibit $A(T) \propto \sqrt{T}$ behavior (see SI). However, there is no generic class of systems in which the prefactor is strictly linear in $T$, and the use of Eyring linearity is therefore not generally justified. Relying on Eyring linearity may even obscure unusual reaction mechanisms. For instance, if the second-order term is dominant, the Arrhenius plot may already show statistically significant deviations from linearity that motivate further investigation, while the corresponding Eyring plot may still appear linear and therefore misleadingly consistent with expectations.

I ended my analysis by proposing a different convention that resolves the dimensionality mismatch between transition and reactant states, leading to a modified activation free energy $\Delta G^\ddagger_{\rm ref}$ based on a freely chosen reference frequency $\nu_{\rm ref}$, for example a typical vibrational frequency or, to remain close to Eyring, $\nu_{\rm ref} = k_B T_{\rm ref}/h$ with fixed $T_{\rm ref}$. In this framework, $\Delta G^\ddagger_{\rm ref}$ can be interpreted as the ratio of probabilities between the reactant state and an augmented transition-state dividing surface with an additional curvature along the orthogonal degree of freedom. This leads to a transparent definition of activation free energy and Arrhenius-based activation enthalpy $\Delta H^\ddagger_{\rm ref} = E_a$ and entropy $\Delta S^\ddagger_{\rm ref} = k_B \ln(A/\nu_{\rm ref})$. Here, $\Delta H^\ddagger_{\rm ref}$ is independent of $\nu_{\rm ref}$ and $\Delta S^\ddagger_{\rm ref}$ changes only by a trivial additive shift. 

This avoids artifacts such as enthalpy–entropy compensation and apparent temperature-dependent activation enthalpies, which arise mainly from the dimensionality mismatch in the Eyring representation rather than physical effects. In particular, for small barriers ($V^\ddagger < 1 k_B T$), the Eyring representation can even predict negative activation enthalpies.
While the Eyring equation itself is correct, there is no general theoretical basis for the expectation that the Eyring plot should be linear. It is therefore more consistent to extract activation parameters from the Arrhenius representation, regardless of whether one uses the Eyring parameter pair $(\Delta H^\ddagger, \Delta S^\ddagger)$ or the reference-frequency-based pair $(\Delta H^\ddagger_{\rm ref}, \Delta S^\ddagger_{\rm ref})$.

\begin{acknowledgement}
I thank 
the Research Council of Norway (project no.~353364) and the COSY Gemini Centre for support. I thank Massimo Bocus for many stimulating discussions and Daan Frenkel for insightful comments on the Gibbs--Helmholtz equation that helped motivate the thermodynamically consistent split presented here. I also thank S\'ebastien van Erp for assistance with figure preparation.
\end{acknowledgement}

\begin{suppinfo}
The Supporting Information contains detailed derivations of the expressions used in the main text.
\end{suppinfo}

\bibliography{biblio}
\end{document}


\section{Gaussian integrals}
\label{sec:si:gaussian}

Below we list standard Gaussian integrals in a form that is particularly useful in statistical thermodynamics, for example $a = k_x$ for $u = x$, $a = 1/m$ for $u = p$, and $a = m$ for $u = v$.
\begin{align}
\label{SI:eq:eq1}
\int_{-\infty}^{\infty} e^{-\beta \frac{1}{2} a u^2}\,du
&= \sqrt{\frac{2\pi}{\beta a}}, \\[6pt]
\int_{0}^{\infty} u\, e^{-\beta \frac{1}{2} a u^2}\,du
&= \frac{1}{\beta a}, \\[6pt]
\int_{-\infty}^{\infty} \frac{1}{2} a u^2 \, e^{-\beta \frac{1}{2} a u^2}\,du
&= 
\frac{1}{\beta}\sqrt{\frac{\pi}{2a\beta}}
\label{SI:eq:eq3}
\end{align}

\section{Derivation of Chandler's reactive flux}
\label{sec:si:extra}
To make the definition of the rate constant precise, we consider an ensemble of phase-space points distributed according to the Boltzmann distribution. This ensemble may be viewed as an arbitrarily large (formally infinite) collection of independent realizations of the system in phase space, each evolving under the same Hamiltonian dynamics.

The first-order rate constant is defined as the short-time limit of the conditional probability for a transition from state $A$ to state $B$,
\begin{equation}
k_{AB} = \lim_{dt \to 0} \frac{1}{dt} \, P(A \to B \ \text{in } dt \mid A \text{ at time } t),
\end{equation}
where $t$ is arbitrary and, without loss of generality, will be set to zero due to time-translational invariance. 
Hence, the rate $k_{AB}$ denotes $(1/dt)$ times the fraction of ensemble members that are in state $A$ at time $t=0$ and in state $B$ at time $t=dt$.

Since being in $A$ or $B$ is decided by the reaction coordinate $q$ being $< q^\ddagger$ or $> q^\ddagger$, we can write this as the ratio of two ensemble averages:
\begin{align}
    k_{AB} = \lim_{dt \to 0} \frac{1}{dt}
    \frac{
    \left \langle
    \theta\left(q^\ddagger-q(0)\right)
    \theta\left(q(dt)-q^\ddagger\right)
    \right \rangle
    }{
    \left \langle 
    \theta\left(q^\ddagger-q\right)
    \right \rangle
    }
    \label{eq:kabtt}
\end{align}
where $q(0)$ and $q(dt)$ denote the values of the reaction coordinate at times $t=0$ and $t=dt$, respectively, and $\theta(x)$ is the Heaviside step function, defined as $\theta(x)=1$ for $x>0$ and $\theta(x)=0$ otherwise. The ensemble average implies an integration over phase space at $t=0$. The denominator does not require an explicit time dependence of $q$ due to time-translation invariance of the equilibrium distribution discussed above.

If we consider the numerator, we see that only a vanishingly small fraction of phase points contributes to the ensemble average in the limit $dt \to 0$, namely those with $q$ on the left side of the dividing surface but sufficiently close to $q^\ddagger$, and with a positive velocity $v$ large enough to satisfy $v\,dt > q^\ddagger - q$.

On the basis of this observation, the following identity can be established \cite{vanErp2012}:
\begin{align}
   \lim_{dt \to 0} \frac{1}{dt}
    \theta\!\left(q^\ddagger-q(0)\right)
    \theta\!\left(q(dt)-q^\ddagger\right)
    = v(0)\, \delta\!\left(q(0)-q^\ddagger\right)\, \theta\!\left(v(0)\right)
\end{align}

Upon substitution of this expression into Eq.~\ref{eq:kabtt}, the explicit time dependence can again be dropped, yielding
\begin{align}
    k_{AB} = 
    \frac{
    \left \langle v\, \delta\!\left(q-q^\ddagger\right)\, \theta\!\left(v\right)
    \right \rangle
    }{
    \left \langle 
    \theta\!\left(q^\ddagger-q\right)
    \right \rangle
    }
    \label{eq:kabvdt}
\end{align}

Evaluating the ensemble averages as integrals over phase space, all momentum (or velocity) degrees of freedom cancel between numerator and denominator, except for the velocity component along the reaction coordinate. We can therefore write
\begin{align}
    k_{AB} = 
    \frac{
    \left(\int e^{-\beta V(q^\ddagger,{\bf q^\perp})} 
    d{\bf q^\perp}\right)
    \left(\int_0^\infty v \, e^{-\beta \frac{1}{2} m_q v^2} \, dv\right)
    }{
    \left(\int_{-\infty}^{q^\ddagger} dq
    \int d{\bf q^\perp}
    e^{-\beta V(q,{\bf q^\perp})} \right)
    \left(\int_{-\infty}^\infty  e^{-\beta \frac{1}{2} m_q v^2} \, dv\right)
    }
    =
    \frac{\int e^{-\beta V(q^\ddagger,{\bf q^\perp})} 
    d{\bf q^\perp}
    }{
    \int_A 
    e^{-\beta V({\bf q})} 
    d{\bf q}} 
    \sqrt{\frac{k_B T}{2 \pi m_q}}
    \label{eq:kab_end}
\end{align}

where we used Eqs.~S1 and S2. 
This expression is identical to Eq.~5 of the main text, thereby proving the reactive flux expression.

As discussed in the main text, the use of nonlinear coordinates complicates the derivation because it requires introducing the Jacobian $|J|$ associated with the coordinate transformation, as well as an effective mass $\mu_q$ for the reaction coordinate $q$, whose units generally differ from kg. In this case, Eq.~\ref{eq:kabvdt} can be written in the more general form\cite{Berne1988, Carter1989}
\begin{align}
    k_{AB}
    &=
    \frac{
    \left\langle
    \delta\!\left(q-q^\ddagger\right)\,
    \theta\!\left(v\right)
    \right\rangle
    }{
    \left\langle
    \theta\!\left(q^\ddagger-q\right)
    \right\rangle
    }
    \sqrt{\frac{k_B T}{2 \pi \mu_q}}
    \nonumber\\
    &=
    \frac{
    \int
    |J(q^\ddagger,\mathbf{q}^\perp)|
    e^{-\beta V(q^\ddagger,\mathbf{q}^\perp)}
    \,d\mathbf{q}^\perp
    }{
    \int_A
    |J(\mathbf{q})|
    e^{-\beta V(\mathbf{q})}
    \,d\mathbf{q}
    }
    \sqrt{\frac{k_B T}{2 \pi \mu_q}},
    \label{eq:kab_endJ}
\end{align}
where the system consists of $N$ particles with Cartesian coordinates $\mathbf{r} = \{r_{i,\alpha}\}$, with $i=1,\dots,N$ and $\alpha \in \{x,y,z\}$. The generalized mass is defined as
\begin{align}
   \mu_q^{-1}
   =
   \left\langle
   \sum_i \frac{1}{m_i}
   \sum_{\alpha=x,y,z}
   \left(
   \frac{\partial q}{\partial r_{i,\alpha}}
   \right)^2
   \right\rangle^\ddagger,
\end{align}
where $\langle \cdots \rangle^\ddagger$ denotes an equilibrium ensemble average constrained to the dividing surface.

\section{Equivalence of QM and Classical Partition Functions}

The classical partition function of a one-dimensional harmonic oscillator, with potential $V(x)=\frac{1}{2}kx^2$ and mass $m$, is given by
\begin{align}
Q_{\rm classical}
&=
\frac{1}{h}
\int_{-\infty}^{\infty}
e^{-\beta \frac{1}{2} k x^2}\, dx
\int_{-\infty}^{\infty}
e^{-\beta \frac{p^2}{2m}}\, dp
\nonumber \\
&=
\frac{1}{h}
\sqrt{\frac{2 \pi k_B T}{k}}
\sqrt{2 \pi m k_B T}
=
\frac{k_B T}{h \nu},
\end{align}
where
$\nu = (1/2\pi) \sqrt{k/m}$ 
is the harmonic frequency. Here, Eq.~S1 was used twice to evaluate the Gaussian integrals over $x$ and $p$.

In quantum mechanics, the energy levels of the harmonic oscillator are given by
\[
E_n = \left(n + \frac{1}{2}\right) h \nu.
\]
The corresponding partition function is therefore
\begin{align}
Q_{\rm QM} &= \sum_{n=0}^{\infty} e^{-\beta E_n} = e^{-\beta \frac{1}{2} h\nu} \sum_{n=0}^{\infty} e^{-\beta n h\nu} = \frac{e^{-\beta \frac{1}{2} h\nu}}{1 - e^{-\beta h\nu}}.
\end{align}
Here we used the infinite geometric series identity $\sum_{n=0}^{\infty} \xi^n = \frac{1}{1-\xi}$ for $|\xi|<1$, with $\xi = e^{-\beta h\nu}$.

To recover the classical limit, we consider $\beta h \nu \ll 1$, corresponding to $h \to 0$. Expanding numerator and denominator in first order and keeping the leading terms  gives
\begin{align}
Q_{\rm QM}=
\frac{e^{-\beta \frac{1}{2} h\nu}}{1 - e^{-\beta h\nu}}
\approx \frac{1 - \beta \frac{1}{2} h\nu}{1 - (1 - \beta h\nu)}
\approx
\frac{1}{\beta h\nu}
= \frac{k_B T}{h\nu}.
\end{align}
We thus arrive at an expression identical to $Q_{\rm classical}$ in Eq.~S9. This shows that the classical limit of the quantum-mechanical partition function reproduces the classical partition function for the harmonic oscillator when using the $1/h^{3N}$ normalization.

This convention for normalizing the classical partition function by $1/h^{3N}$ ensures that the quantum-mechanical and classical partition functions coincide in the classical limit $h \rightarrow 0$. It is important, however, to emphasize that the Boltzmann weight only determines relative probabilities: the probability of a microstate can be written as $\mathrm{const.}\, e^{-\beta H}$, where the choice of constant fixes the overall magnitude of the partition function. In quantum mechanics, choosing a constant different from unity is unusual but possible. In classical mechanics, taking the constant equal to one yields the standard phase-space partition function, typically denoted $Z$, which satisfies $Z = h^{3N} Q$ and contains no explicit dependence on Planck’s constant, whereas choosing the constant as $h^{-3N}$ leads to $Q$. It is important to stress that this distinction is purely conventional and not physically fundamental.

\section{Harmonic approximation in 2D}

We consider a two-dimensional system with coordinates $x$ and $y$, described by the potential energy surface $V(x,y)$ shown in Fig.~1 of the main text, where $y$ denotes the coordinate orthogonal to the reaction coordinate $x$.
For this model, the integrals occuring in Eq.~5 of the main article can be expresse as:

\begin{align}
\int e^{-\beta V(q^\ddagger, {\bf q^\perp})} d{\bf q^\perp}&=
\int_{-\infty}^\infty e^{-\beta V(x^\ddagger, y)} dy \nonumber \\
\int_A e^{-\beta V({\bf q})} d{\bf q}&=
\int_{-\infty}^{x^\ddagger} dx \, \int_{-\infty}^\infty  dy \, e^{-\beta V(x, y)} dy
\label{eq:S12}
\end{align}

 We assume that near the minimum of the reactant well and near the transition state, the potential can be approximated as
\begin{eqnarray}
V(x,y) &\approx& \frac{1}{2} k_{x}^R x^2 + \frac{1}{2} k_{y}^R y^2 \quad \text{for } x \approx 0,\; y \approx 0, \nonumber \\
V(x^{\ddagger},y) &\approx& V^{\ddagger} + \frac{1}{2} k_y^{\ddagger} y^2 \quad \text{for } y \approx 0.
\end{eqnarray}
Since the integrands are maximal near the minima of the potentials and decrease rapidly away from these regions, the dominant contribution to the configurational integrals arises from the region where the quadratic expansion of the potential is valid.
The integrals of Eq.~\ref{eq:S12}
are hence evaluate as
\begin{align}
&\int_{-\infty}^\infty e^{-\beta V(x^\ddagger, y)} dy \approx
\int_{-\infty}^\infty e^{-\beta (
V^{\ddagger} + \frac{1}{2} k_y^{\ddagger} y^2)
} dy=
e^{-\beta 
V^{\ddagger}}
\sqrt{\frac{2\pi k_B T}{k_y^\ddagger}}
 \\
&
\int_{-\infty}^{x^\ddagger} dx \, \int_{-\infty}^\infty  dy \, e^{-\beta V(x, y)} 
\approx 
\int_{-\infty}^{\infty} dx \, \int_{-\infty}^\infty  dy \, e^{-\beta (\frac{1}{2} k_{x}^R x^2 + \frac{1}{2} k_{y}^R y^2)} 
=\sqrt{\frac{2\pi k_B T}{k_x^A}}
\sqrt{\frac{2\pi k_B T}{k_y^A}} \nonumber
\end{align}
where Eq.~\ref{SI:eq:eq1} has been applied three times and Eq.~5 of the main text simplifies to:
\begin{align}
k_{AB} &=
\sqrt{\frac{k_B T}{2 \pi m}}
\left(
\frac{
e^{-\beta 
V^{\ddagger}}
\sqrt{\frac{2\pi k_B T}{k_y^\ddagger}}
}{
\sqrt{\frac{2\pi k_B T}{k_x^A}}
\sqrt{\frac{2\pi k_B T}{k_y^A}} 
}
\right)=
\frac{1}{2 \pi}\sqrt{\frac{k_x^A}{m}} 
\left(
\frac{
\sqrt{k_y^A}
}{
\sqrt{k_y^\ddagger} 
}
\right)e^{-\beta 
V^{\ddagger}}
= \frac{\nu_x^A \nu^A_y}{\nu^\ddagger_y} e^{-\beta 
V^{\ddagger}}
\end{align}
with $\nu_\alpha^X = \frac{1}{2\pi}\sqrt{k_\alpha^X/m_\alpha}$ the characteristic frequency along coordinate $\alpha$ in state $X$. This expression can be directly generalized to higher dimensions. For a system with $3N$ degrees of freedom, one obtains
\begin{equation}
k_{AB} = 
\frac{\prod\limits_{i=1}^{3N} \nu_i^{A}}
{\prod\limits_{j=1}^{3N-1} \nu_j^{\ddagger}}
\, \exp\left(-\beta V^\ddagger\right),
\end{equation}
which is the well-known result of harmonic transition state theory (HTST), where the single unstable mode at the transition state is excluded from the product.

\section{On the enthalpy--entropy compensation effect}

The enthalpy--entropy compensation effect suggests a linear relationship between $\Delta H^\ddagger$
and $\Delta S^\ddagger$ of the form
\begin{align}
\Delta H^\ddagger = a + b\, \Delta S^\ddagger.
\end{align}

Let us consider Eqs.~14 of the main article:
\begin{align}
\Delta H^\ddagger(T) &= V^\ddagger - k_B T, \quad
\Delta S^\ddagger(T) = k_B \left[ \ln \left( 
\frac{h \nu_x^A \nu_y^A}{k_B T \nu_y^\ddagger} 
 \right) - 1 \right].
\label{eq:right_split}
\end{align}

Consider a set of experiments where the rate of a process is measured over a limited temperature range, and let $T'$ denote the midpoint of this range such that $T = T' + \delta T$.

Expanding $\Delta S^\ddagger(T)$ around $T'$ gives
\begin{align}
\nonumber
\Delta S^\ddagger(T)
&= k_B \left[ \ln \left( 
\frac{h \nu_x^A \nu_y^A}{k_B (T' + \delta T)\, \nu_y^\ddagger} 
 \right) - 1 \right] 
\approx
k_B \left[ \ln \left( 
\frac{h \nu_x^A \nu_y^A}{k_B T' \nu_y^\ddagger} 
 \right) - \frac{\delta T}{T'} - 1 \right] 
 \\&
 =
k_B \ln \left( 
\frac{h \nu_x^A \nu_y^A}{k_B T' \nu_y^\ddagger} 
 \right)
- \frac{k_B}{T'}(T - T')-k_B
=
k_B \ln \left( 
\frac{h \nu_x^A \nu_y^A}{k_B T' \nu_y^\ddagger} 
 \right)
-\frac{k_BT}{T'}.
\end{align}
which yields  $\Delta H^\ddagger \approx a + b\, \Delta S^\ddagger$ with paramters
\begin{align}
a &= V^\ddagger - k_B T' \ln \left( 
\frac{h \nu_x^A \nu_y^A}{k_B T' \nu_y^\ddagger} 
 \right), \nonumber \\
b &= T'.
\end{align}

\section{Hard-core models}

Hard-core models can be viewed as an idealized limit of strong anharmonicity, but they are nevertheless relevant for a class of physically realistic systems. In these models, the potential energy (or free-energy) surface is nearly flat over large regions of configuration space and changes rapidly only in a narrow transition region.

One example is gas-phase reactions, where the potential energy remains approximately constant until a steep increase occurs upon collision. Similar behavior can also arise in solvated systems, where the free-energy surface of the reactant state is relatively flat over an extended region before rising sharply near the transition state region.

\begin{figure}[h]
    \centering
    \includegraphics[width=0.6\textwidth]{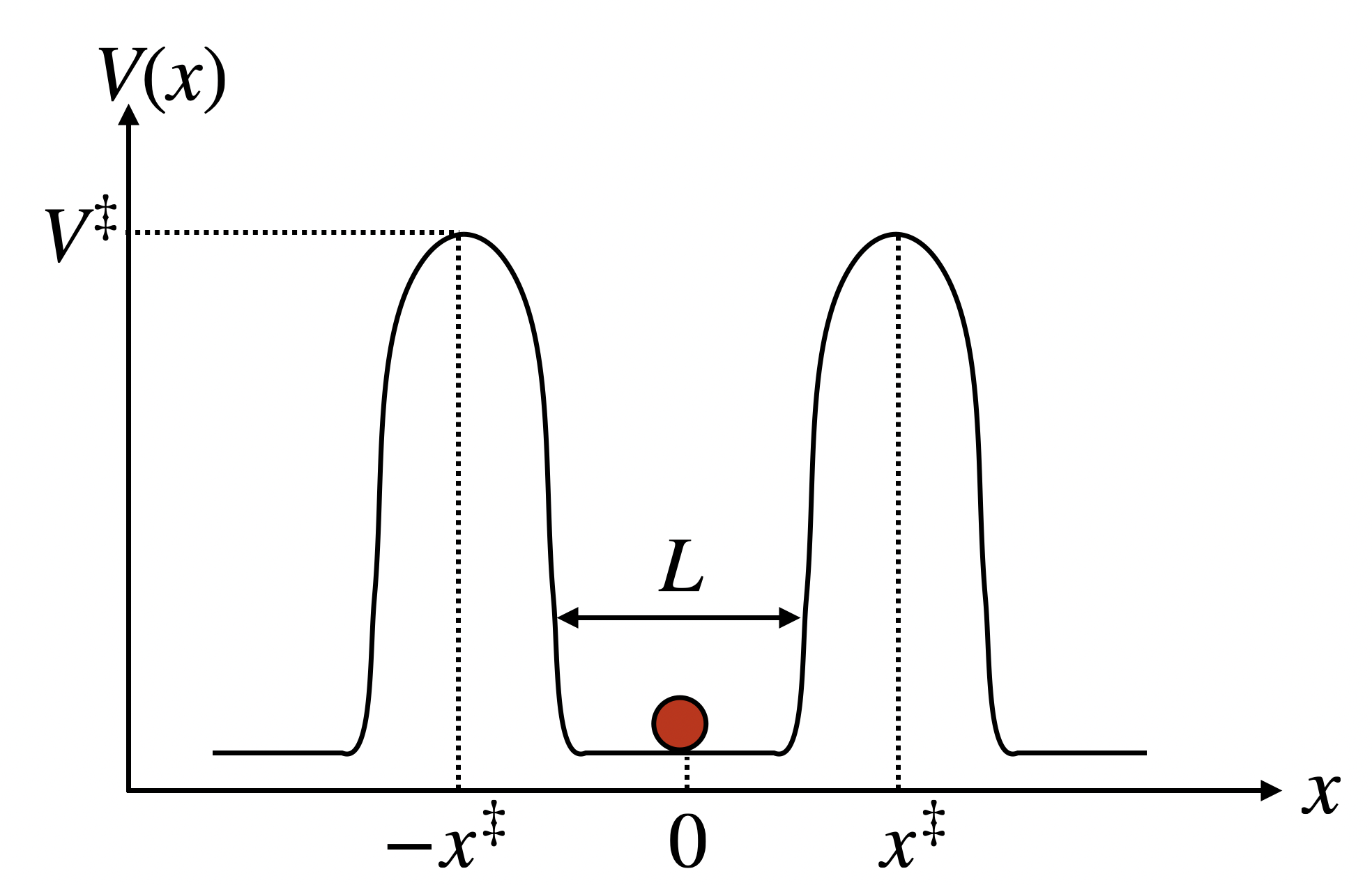}
    \caption{Schematic illustration of a particle confined in a water slab between two membranes, representing a system that can exhibit hard-core-like free-energy behavior.}
    \label{fig:membrane}
\end{figure}

A representative example is permeation through a membrane, such as a particle confined in a water slab between two membranes (Figure~\ref{fig:membrane}).

Here, we consider a one-dimensional reaction coordinate $x$ with a corresponding potential energy surface (PES) $V(x)$. In this simple case, the free energy and potential energy are identical, i.e., $G(x)=V(x)$.

The denominator in Chandler's reactive flux expression (Eq.~1) is then

\begin{align}
\int_A e^{-\beta G(q)}\,dq
=
\int_{-x^\ddagger}^{x^\ddagger} e^{-\beta V(x)}\,dx
\approx L,
\end{align}

where $L$ is the width of the nearly flat reactant region. 
The rate constant for escape from the confinement is therefore given by
\begin{align}
k
=
2\sqrt{\frac{k_B T}{2\pi m}}
\frac{e^{-\beta V^\ddagger}}{L}.
\end{align}

The factor of 2 arises because the particle can escape from the reactant region through either the left or the right confining boundary.
Interestingly, the factor $\sqrt{k_B T/(2\pi m)}$ in Chandler's expression remains fully recognizable after evaluating the free-energy terms in this hard-core model. This contrasts with the factor $k_B T/h$ in the Eyring equation, which does not reflect the underlying trends with mass $m$ or temperature $T$.

If the reactant-state free energy is expected to be flat, as in the above example, the hypothetical transition-state potential can be adapted straightforwardly. Instead of the harmonic restraint $V^\ddagger_{\mathrm{ref}}(q,\mathbf{q}^\perp)=V(q^\ddagger,\mathbf{q}^\perp)+\tfrac{1}{2}k_{\mathrm{ref}}(q-q^\ddagger)^2$, one may employ an infinite square-well restraint, $V^\ddagger_{\mathrm{ref}}(q,\mathbf{q}^\perp)=V(q^\ddagger,\mathbf{q}^\perp)+B(q,L_{\rm ref})$, where $B(q,L_{\rm ref})=0$ for $|q-q^\ddagger|<L_{\rm ref}/2$ and $B(q,L_{\rm ref})=\infty$ otherwise. The derivation then proceeds analogously to Eq.~22 of the main article, with $\sqrt{2\pi k_B T/k_{\rm ref}}$ replaced by $L_{\rm ref}$. The latter is then a hypothetical mass-dependent reference length proportional to $1/\sqrt{m_q}$, thereby eliminating the mass dependence of the prefactor.

To recover an expression closely resembling the Eyring equation, one may choose $L_{\rm ref}=h/\sqrt{2\pi k_B T_{\rm ref}m_q}$, which yields
\begin{align}
k
=
\frac{k_B\sqrt{T_{\rm ref}T}}{h}
\exp\!\left(-\beta\Delta G^\ddagger_{\rm ref}\right).
\end{align}

In this formulation, only a fixed reference temperature $T_{\rm ref}$ is required to define $\Delta G^\ddagger_{\rm ref}$. Consequently, a linear relation is expected between $\ln\!\bigl(kh/(k_B\sqrt{T_{\rm ref}T})\bigr)$ and $1/T$, from which $\Delta H^\ddagger_{\rm ref}$ and $\Delta S^\ddagger_{\rm ref}$ can be obtained from the slope and intercept, respectively. These quantities enter exactly as on the right-hand side of Eq.~23, although their thermodynamic definitions differ slightly. In particular, in $\Delta G^\ddagger_{\rm ref}$ in Eq.~24, $\sqrt{2 \pi k_B T/k_{\rm ref}}$ is replaced by $L_{\rm ref}$, while in $\Delta H^\ddagger_{\rm ref}$ in Eq.~25 the $\tfrac{1}{2}k_B T$ term is omitted.

The entropy of activation follows again from Eq.~27, yielding a contribution proportional to $L_{\rm ref}/L$ for a model system such as the one depicted in Fig.~\ref{fig:membrane}. This demonstrates that consistent behavior and a linear logarithmic plot can be obtained, with approximately temperature-independent enthalpies and entropies of activation, by adapting the virtual transition-state potential along the $q$ coordinate to mimic the behavior of the reactant state.
\bibliography{biblio}